# SOIL MOISTURE MONITORIZATION USING GNSS REFLECTED SIGNALS

A. Egido, G. Ruffini, M. Caparrini, C. Martín, E. Farrés, X. Banqué

*Starlab Barcelona*
*Edifici de l'Observatori Fabra, Muntanya del Tibidabo*
*Camí de l'Observatori Fabra s/n - 08035 – Barcelona - Spain*
*Email: alejandro.egido@starlab.es*

## 1. INTRODUCTION

The use of GNSS signals as a source of opportunity for remote sensing applications has been a research area of great interest since 1993, when M. Martin Neira (ESA) proposed that GPS signals reflected from the Earth's surface could be detected to retrieve ocean altimetry information accounting for the existent delay between the direct and the reflected signals (the PARIS concept, PAssive Reflectometry and Interferometry System) [1]. Since then, several applications based on a GNSS bistatic radar configuration have been developed taking advantage of the high availability and stability of GNSS signals. This technique is commonly known as GNSS-R (Global Navigation Satellite System – Reflections).

GNSS-R studies and investigations have been mainly focused on sea surface topography. Within this frame, Starlab Barcelona has developed Oceanpal®, a fully operational system that can provide GNSS-R data and higher level products, such as real time significant wave height (SWH) and altimetry data.

The application of GNSS-R to land remote sensing has been largely overlooked. Nevertheless, there is experimental evidence that GPS reflected signals from the ground can be detected and processed in order to obtain soil moisture estimates, [2][3][4].

The importance of soil moisture relies in the fact that it is a prime parameter for the surface hydrology cycle, which is one of the keys for the understanding of the interaction between continental surfaces and the atmosphere in environmental studies. Water storage in the soil, either in the surface layer or in deeper levels, affects not only the evapotranspiration but also the heat storage ability of the soil, its thermal conductivity, and the partitioning of energy between latent and sensible heat fluxes. In addition, the value of the surface layer volumetric soil moisture direct evaporation from soil, and determines the possibility of surface runoff after rainfalls.

Despite the recognised relevance of soil moisture, providing such parameter on global scales remains a significant challenge. Sensors based on GNSS-R offer this possibility and could represent a very important milestone in the development of a global soil moisture model. Starlab expects to develop an operational GNSS-R sensor oriented to soil moisture retrieval, which will be based in the Oceanpal® instrument's architecture. The present paper reviews the most important theoretical aspects to take into consideration for the development of a GNSS-R soil moisture sensor, and suggests how the use of the forthcoming Galileo signals might help in this task.

## 2. SOIL MOISURE ESTIMATION WITH GNSS SIGNALS

The basis for the retrieval of soil moisture with GNSS-R systems lays in the variability of the ground dielectric properties associated to soil moisture. Higher concentrations of water in the soil yield a higher dielectric constant and reflectivity. Consequently, the reflected signal's peak power can be related to soil moisture.

Previous investigations [2-7] have demonstrated the capability of GPS bistatic scatterometers to obtain signal to noise ratios high enough to sense small changes in surface reflectivity. Furthermore, these systems present some advantages with respect to those currently used to retrieve soil moisture. First, GPS signals lie in L band, which is the most sensitive band for soil moisture microwave remote sensing. Secondly, in contrast to microwave radiometry, variations on thermal background do not dramatically contaminate the GPS reflected signals. As will be seen below, thermal background influences soil moisture observables, but this effect is not as important as for microwave radiometry. Thirdly, GPS scatterometry from space has a potential higher spatial resolution than microwave radiometry, due to the highly stable carrier and code modulations of the incident signals which enables the use of Delay Doppler mapping.

Nevertheless, in order to obtain precise soil moisture estimates there are several phenomena that need to be taken into consideration, mainly the effects of diffuse scattering over the soil surface: soil roughness and vegetation canopy.

## 3. THE OCEANPAL® INSTRUMENT

Oceanpal® is a GNSS-R based sensor designed for operational coastal monitoring. It is an inexpensive, all-weather, dry and passive instrument which can be deployed on multiple platforms, static (coasts, harbours, off-shore), and slowly moving (boats, floating platforms, buoys). In its present form, Oceanpal® can deliver two kinds of Level-2 products: sea-surface height and significant wave height (SWH). However, due to its flexibility in terms of data acquisition, the Oceanpal® instrument can be applied to the retrieval of soil moisture in a straightforward way.

### 3.1. Instrument's Architecture

Oceanpal® comprises three subsystems: a radio frequency (RF) section, an intermediate frequency section and a data processing section. The basic system architecture is illustrated in Fig. 1. The RF section features a pair of low gain L-band antennas. An RHCP (Right-hand circular polarized) zenith antenna collects the direct GNSS signals while an LHCP (left-hand circular polarized) nadir antenna collects the sea-surface reflected GNSS signals. Data bursts of some minutes are acquired from each channel using two radio frequency front-ends that down-convert the signal to intermediate frequency (IF). The acquisition time is a parameter that can be specified by the user.

Within the IF section, the signal is one-bit sampled and stored on a hard disk. After the acquisition process, these direct and reflected raw data are then fed into the processing section of the instrument where a pair of software GNSS receivers detects and tracks the available signals in the direct channel (which works as master) and blindly dispreads the reflected signals in the reflected/slave channel. The result of this processing is a set of direct and reflected electromagnetic field time series (complex waveforms) for each satellite in view, plus some ancillary information. The complex waveforms are then used to produce higher level products by the data processing algorithms.

It must be noted at this point that Oceanpal® is currently a GPS based instrument. However, the interoperability of the GPS and Galileo L1 signals and the fact that Oceanpal® is implemented as a software receiver (after the digitization of the signal) the evolution of the system towards a GPS and Galileo instrument is relatively easy. Starlab Barcelona expects to have this GNSS-R instrument working by the beginning of 2008.

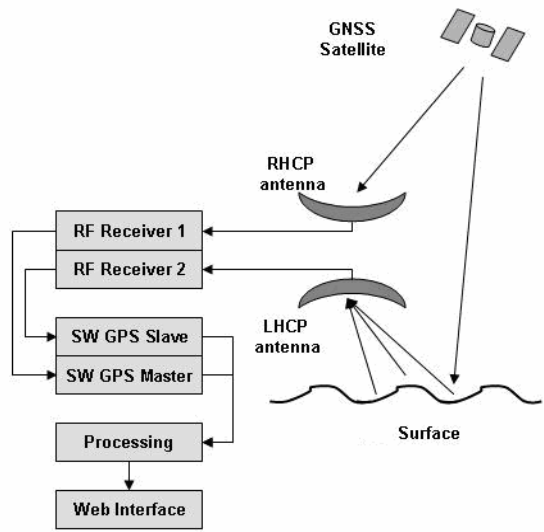

Fig. 1. Oceanpal basic setup.

## 3.2. The Interferometric Complex Field

The fundamental product of the instrument data processing chain is the so called Interferometric Complex Field (ICF), which is a time series calculated as the ratio between the reflected and direct waveform peaks, (1)

$$ICF(t) = \frac{p_R(t)}{p_D(t)} \tag{1}$$

where $p_R$ and $p_D$ represent the time series of waveform peaks for the reflected and direct signals, respectively. The ICF is the basis for the algorithms used to calculate different higher order level products, and represents also the fundamental magnitude for soil moisture estimation.

## 4. SOIL MOISTURE ESTIMATION USING OCEANPAL®

In the previous section, the interferometric complex field has been defined. It must be noted in (1) that the squared absolute value of the ICF can be considered as the peak power ratio of the reflected and the direct signal waveforms in the lapse of time t. This means that it represents a measure of the surface reflectivity, which in turn, as mentioned in section 2, can be related to the soil moisture volumetric content.

However, there are several parameters such as surface roughness, vegetation canopy, and thermal background, which affect the determination of soil moisture. Their effects are reviewed more in depth in the next section. Considering, as a first approximation, that the only parameter affecting the reflected signal is the soil reflectivity, the following incoherent averaging can be performed:

$$\Gamma_{av} = \frac{1}{N} \sum_{i=1}^{N} |ICF(t)|^2 \tag{2}$$

where N is the number of waveforms computed during one data acquisition. With (2) an averaged value of the soil's reflectivity $\Gamma_{av}$ is be obtained for the whole acquisition time span, typically one minute. From the Fresnel equations of reflection, the reflection coefficients for vertical and horizontal polarization can be obtained:

$$\Gamma_v = \frac{\varepsilon_r \sin\gamma - \sqrt{\varepsilon_r - \cos^2\gamma}}{\varepsilon_r \sin\gamma + \sqrt{\varepsilon_r - \cos^2\gamma}} \tag{3}$$

and

$$\Gamma_h = \frac{\sin\gamma - \sqrt{\varepsilon_r - \cos^2\gamma}}{\sin\gamma + \sqrt{\varepsilon_r - \cos^2\gamma}} \tag{4}$$

where $\gamma$ is the incidence angle. The GPS signals is mostly right hand circular polarized (RHCP), which means that it presents a vertical and a horizontal polarization component. However, for high incidence angles, e.g. above 60º, the difference between the reflections coefficients for vertical and horizontal polarization can be considered negligible, in a first approach, and therefore, just the reflection coefficient for vertical polarization can be taken into account. As noted in [4], the error in using just the vertical value is 5% of the reflectivity at the worst incidence angle. Identifying $\varepsilon_r$ with $\varepsilon_{soil}$ and solving the previous equation for the permittivity one can show that

$$\varepsilon_{soil} = \frac{1 \pm \sqrt{1 - 4\sin^2\gamma \cdot \cos^2\gamma \cdot \left(\frac{1-\Gamma}{1+\Gamma}\right)^2}}{2\sin^2\gamma \cdot \left(\frac{1-\Gamma}{1+\Gamma}\right)^2} \tag{5}$$

In order to relate the soil permittivity to soil moisture a semi-empirical model presented in [11] can be used. The authors suggest that the polynomial that describes the relationship between soil moisture and dielectric constant for a frequency around 1.4 GHz is given by

$$\varepsilon_{soil} = 2.862 - 0.012S + 0.001C + (3.803 + 0.462C - 0.341)m_v + (119.003 - 0.500S + 0.633C)m_v^2 \quad (6)$$

where S and C are the sand and clay textural compositions of a soil in percent by weight, and $m_v$ is the volumetric soil moisture. This model has been used as a semi-empirical approach in [5] with acceptable results.

Another partially different approach for determine the complex permittivity of the soil using navigation reflected signals has been proposed in [2][12]. The main difference in this approach is the fact that only one antenna is used to collect both the direct and the reflected signal, thus obtaining an interferometric field as sum of the two EM waves.

Other more complicated and accurate models than those presented above have also been used. For instance, in [6] a model based on Kirchoff Approximation and Geometric Optics is adopted. This model has been successfully applied in modelling various GNSS bistatic radar scenarios (see for instance [7] which is the most commonly used model in GNSS scattering from ocean surfaces).

## 5. CONSTRAINT IN THE DETERMINATION OF SOIL MOISTURE

As mentioned in the previous section, the estimation of soil moisture with L band signals is affected by several ancillary phenomena that distort the interferometric complex field and bias the measurements of the soil reflectivity. The present section reviews the most important ones and their effects in the scattered signals.

### 5.1. Surface Roughness and Vegetation Canopy

Under the assumption of a flat scattering surface, the GPS signal emitted by a certain space vehicle would be reflected basically from a specular point over the surface (more precisely, from the first Fresnel zone [8]). It is known that the specular point is determines the shortest path between the emitter and the receiver, through the reflecting surface. However, if a rough surface is considered and according to the Geometrical Optics model, slopes may exist with the proper orientation to redirect the incoming radiation to the receiver antenna from sites away from the specular point, as depicted in Fig.1. The locus of points over the surface from which the reflected signal arrives at the same delay, with respect to the specular point, is given by ellipses which are called iso-delay lines. The signals will also be affected by Doppler shifts due to the changing geometry of the scenario and the relative motion of emitter and receiver, however this effect will not be considered for the moment for soil moisture retrieval. Note that this approximation is perfectly valid in the case of a fix receiver.

The power signal in the receiver at any delay is the result of summing up the reflected field from each individual scatterer within the corresponding iso-delay ellipse. Assuming that natural scenes are composed of independently phased scatterers, the resulting composite signal is stochastic with Rayleigh distribution (i.e., affected by speckle [8]). Note that the signal strength received from an individual scatterer depends on the reflection coefficient of each surface element, as well as the incidence angle. In addition, further scattering and attenuation occurs when the signal path includes vegetation canopy.

GNSS receivers rely on the spread spectrum properties of the pseudo-random noise code (PRN), which modulates the carriers, in order to track the signals. The tracking is performed through the correlation of the received signals with a clean replica of itself (matched filtering), obtaining a waveform whose shape resembles, in an average sense, the autocorrelation of the original PRN code. However, for the case of the reflected signal, in addition to the delay introduced as a consequence of a longer signal path, the waveform is in general distorted and its peak power diminished due to the scattering process. In Fig.2, ideal waveforms of the direct and reflected signals are sketched.

As mentioned before, the peak power of the reflected signal waveform depends on the surface reflectivity and therefore can be related to soil moisture. However, since the reflected signal becomes a stochastic process through the scattering process, measuring the peak of the waveform is not a straightforward task.

The effects of vegetation canopy in the scattering process of GNSS signals are a very important factor to be taken into consideration [4]. Specifically, for soil moisture remote sensing, vegetation is often modelled separately from the bare soil surface as a signal attenuation which is proportional to vegetation water content [9].

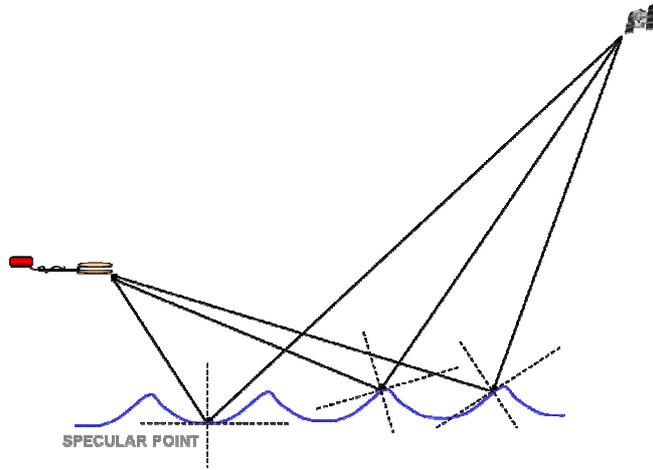

Fig.1. GPS reflections over a rough surface. Scatterers outside the specular point exist that have the proper inclination to redirect the incident signal towards the receiver.

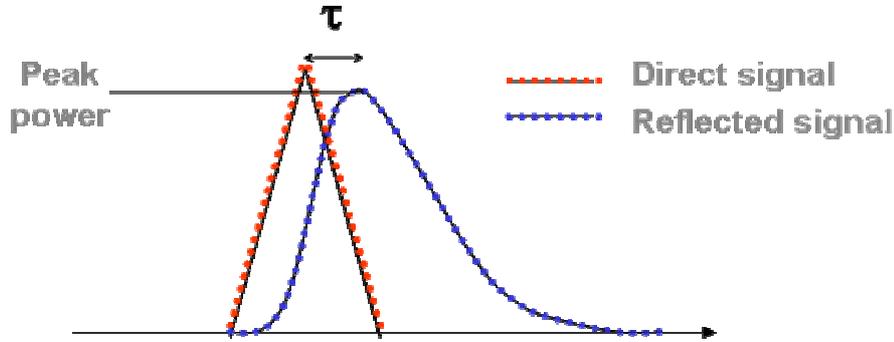

Fig.2. Direct and reflected signal waveforms

**5.3. System Noise**

Back to the assumption of a perfectly smooth reflecting surface, the only changes affecting in the reflected signal with respect to the direct one would be the amplitude, decreased by the reflectivity of the surface, and a phase shift. Thus, the peak power of the reflected signal normalized by the peak power of the direct signal provides an observable that is proportional to the soil's reflectivity. However, the only direct observable of GPS signals is the waveform that results from the cross-correlation of the incoming signal with the locally generated PRN code, as explained above.

It can be demonstrated, although it is not within the scope of this review document, that the waveform's peak powers of the direct and reflected signals are proportional to the signal to noise ratio of the received signal. In the realistic case of a rough scattering surface, the reflected signal is affected by additive Gaussian white noise of thermal origin, plus speckle. As a consequence of the speckle noise, the waveform shape is distorted and the peak power fluctuates due to fading effects. Thus, estimation of the peak power requires a certain amount of incoherent averaging in order to reduce uncertainty in the measurements. The equation that models the mean complex waveform power can be approximated as

$$\left\langle \left| \hat{C}_k(\tau) \right|^2 \right\rangle = s^2 \Lambda^2(\tau) + f(T, \tau_n) \tag{7}$$

where $s$ is the signal SNR, $\Lambda(\tau)$ indicates the triangle correlation function of the PRN code, and $f$ is a term which is function of the coherent integration time and the noise coherence time. The caret denotes independent stochastic variables that have some probability distribution function (a combination of Gaussian thermal noise and Rayleigh speckle). The lag variable $\tau$ identifies different variables. It can be inferred from the previous relation that the soil moisture observable will be affected by the variations of thermal background through variations in the signal to noise ratio, which will need to be accounted for in the inversion process for soil moisture estimates.

Another significant effect of the system noise is the maximum allowed precision with which the waveform peak power can be determined. The minimum variance associated with an estimator is defined by the Cramer-Rao lower bound. The Cramer-Rao lower bound for estimating a peak power $\alpha_p^2$ is given by [10]

$$\alpha^2_{\alpha_p} = \frac{\alpha_p^2}{N\Upsilon} \qquad (8)$$

where N is the number of independent samples averaged and $\Upsilon$ is the detected signal energy to noise ratio. In order to increase the signal energy to noise ratio, longer integration times should be used when performing the correlation of the incoming signal with the clean replica. However, the coherent integration time has a natural practical upper limit, for the case of GPS signals, which corresponds to the duration of a navigation bit, e.g., 20ms for GPS C/A. In the case of Galileo signals, the existence of pilot signals which are not modulated by a navigation message eliminates the upper limit restriction for the integration time. Therefore the variance of the waveform peak power estimation should be significantly reduced by using this new signal.

The other way to further reduce the uncertainty in the final estimation is by averaging independent samples. Independency of observations implies that the phases of the scatterers are uncorrelated in each data set to be averaged. This independency translates into a sufficient change of the geometry of the reflection in subsequent data takes. Considering a ground based instrument, since the soil surface is static, the independency condition is accomplished by allowing enough time to pass between consecutive samples, so that the geometry of the observations is sufficiently different because of the movement of the transmitting GPS space vehicles. The fact that a GNSS-R based instrument (such as Oceanpal®) is capable to simultaneously track and process signals from several satellites, provides additional independent measures of the scattering surface, which will also contribute to reduce the variance of the estimation.

It is important to stress that in the approach to soil moisture estimation presented, variations in the ICF which can be caused by the different thermal background fluctuations in the direct and reflected signals, as well as any mismatching in the direct and reflected receiving chains have to be accounted for since they directly impact the magnitude used for the estimation. Similarly, since most GNSS-R systems use one zenith and one nadir antenna, a temperature gradient in the system is likely to occur, which can result in an additional difference of the noise affecting the direct and the reflected receiving chains. Hence, both receiving chains will need to be calibrated to perform effective soil moisture estimation.

## 6. CONCLUSIONS

The most important aspects and constraints for soil moisture retrieval with a GNSS-R based instrument, as well as a simple first-approach scattering model have been reviewed in this article. In addition, a GPS-R instrument by Starlab (Oceanpal®), suitable for soil moisture retrieval, has been presented. The state of the art does not provide the capability to perform GNSS-R soil moisture remote sensing in an accurate and precise way. Further investigations are needed to relate in a more precise way the effects of soil moisture to the GNSS bistatic scattering process; differences between vertical and horizontal reflection coefficients will need to be accounted for, as well as diffuse scattering effects due to surface roughness, adverse effects caused by temperature variations, and vegetation canopy should also be included in forthcoming scattering models. Extensive validation and calibration campaigns will need to be performed, so that soil moisture estimates can be compared and related to in-situ soil moisture measurements with precise knowledge of surface roughness and vegetation canopy.

Concerning the use of GALILEO, whereas for other remote sensing applications it has been proven to represent a big step forward (for example in ocean mesoscale altimetry [13]), with respect to soil moisture estimation, at the moment, the specific contribution of these new signals seems to be limited to the possibility of longer integration time and longer codes, for noise reduction, and to the use of more frequencies and satellites to increase the number of measurements.

Notwithstanding the highlighted difficulties, soil moisture remote sensing with GNSS-R remains an interesting goal, taking into account the huge advance it would represent in obtaining global estimations of such an important parameter for the hydrologic cycle.